\documentclass[prd,showpacs,twocolumn]{revtex4-1} 
\usepackage{graphicx}
\usepackage{amsmath}
\usepackage{amssymb}
\usepackage{bm}

\newcommand{\nn}{\nonumber}

\begin{document}
%\subheader{draft   ~~~\today}

\title{Exact Holography of the Mass-deformed M2-brane Theory }
\author{Dongmin Jang}
\email{dongmin@skku.edu}
\affiliation{Department of Physics,~BK21 Physics Research Division,
~Institute of Basic Science, Sungkyunkwan University, Suwon 440-746, South Korea}
\author{Yoonbai Kim}
\email{yoonbai@skku.edu}
\affiliation{Department of Physics,~BK21 Physics Research Division,
~Institute of Basic Science, Sungkyunkwan University, Suwon 440-746, South Korea}
\author{O-Kab Kwon}
\email{okab@skku.edu}
\affiliation{Department of Physics,~BK21 Physics Research Division,
~Institute of Basic Science, Sungkyunkwan University, Suwon 440-746, South Korea}
\author{D.~D. Tolla}
\email{ddtolla@skku.edu}
%\affiliation{Department of Physics,~BK21 Physics Research Division,
%~Institute of Basic Science, Sungkyunkwan University, Suwon 440-746, %South Korea}
\affiliation{Department of Physics,~BK21 Physics Research Division,
~Institute of Basic Science, and  University College,
Sungkyunkwan University, Suwon 440-746, South Korea}
%\date{\today}

\begin{abstract}

We test the holographic relation between the vacuum expectation 
values of gauge invariant operators 
in ${\cal N} = 6$ U$_k(N)\times {\rm U}_{-k}(N)$ 
mass-deformed ABJM theory  
and the LLM  geometries 
with $\mathbb{Z}_k$ orbifold in 11-dimensional supergravity.
To do that, we apply the Kaluza-Klein reduction 
to construct a 4-dimensional gravity theory and 
implement the holographic renormalization procedure.
We obtain an exact holographic relation 
for the vacuum expectation 
values of the chiral primary operator with conformal 
dimension $\Delta = 1$, which is given by $\langle {\cal O}^{(\Delta=1)}\rangle= N^{\frac32} \, f_{(\Delta=1)}$, for large $N$ and $k=1$. 
Here factor $f_{(\Delta)}$ is independent of $N$.
Our results involve infinite number of exact dual relations 
for all possible supersymmetric Higgs vacua 
and so provide a nontrivial test of gauge/gravity duality 
away from the conformal fixed point.  
We also extend our results to the case of  $k\ne 1$ for LLM geometries represented by 
rectangular-shaped Young-diagrams. 

\end{abstract}
%\keywords{gauge/gravity duality, mass-deformed ABJM theory, 
%LLM geometry}
%\pacs{11.25.Tq, 11.27.+d, 03.65.Ud}

%\newpage
%\maketitle
%\today\\
\maketitle

\section{Introduction}
%{\it Introduction.}---
In the context of the AdS/CFT 
correspondence~\cite{Maldacena:1997re,Gubser:1998bc,Witten:1998qj}, 
it was conjectured that 
the string/M theory on AdS$_{d+1}\times {\cal X}$ with a compact manifold 
${\cal X}$ is dual to 
$d$-dimensional conformal field theory (CFT). 
The conjecture was soon extended to quantum field theories (QFTs) 
which can be obtained from the CFTs at the 
Ultraviolet (UV) fixed point by adding relevant operators 
to the action  or considering vacua where the conformal 
symmetry is broken. 
Then the dual geometries for those QFTs are asymptotic to 
AdS$_{d+1}\times {\cal X}$. 
Due to computational difficulties on both sides,  
most of the efforts to test the duality have been
focused on the large $N$ limit of the QFT, $N$ being 
the rank of the gauge group. 

In this letter, we analyze a model which shows 
a supporting evidence for an exact dual relation away from the conformal fixed point 
in the large $N$ limit. 
We consider the ${\cal N}=6$ ${\rm U}_{k}(N)\times {\rm U}_{-k}(N)$ 
Aharony-Bergman-Jafferis-Maldacena (ABJM) theory with Chern-Simons level 
$k$~\cite{Aharony:2008ug}, as the CFT at the UV fixed point. 
The ABJM theory allows the supersymmetry preserving 
mass deformation  and the deformed 
theory (mABJM)~\cite{Hosomichi:2008jb,Gomis:2008vc} 
has discrete Higgs vacua presented by the Gomis, Rodriguez-Gomez, Van Raamsdonk, Verlinde (GRVV) matrices~\cite{Gomis:2008vc}. 
It was known that the vacua of the mABJM theory 
have one-to-one correspondence~\cite{Kim:2010mr,Cheon:2011gv} with the half BPS 
Lin-Lunin-Maldacena (LLM) geometries~\cite{Bena:2004jw, Lin:2004nb} 
with ${\mathbb Z}_k$ orbifold having SO(2,1)$\times$SO(4)/${\mathbb Z}_k\times$SO(4)/${\mathbb Z}_k$ 
isometry in 11-dimensions~\cite{Auzzi:2009es,Cheon:2011gv}. 
Since the mABJM theory is obtained by a relevant deformation 
from the ABJM theory at the UV fixed point, 
the dual geometry should be asymptotically 
AdS$_4\times S^7/{\mathbb Z}_k$.

Here we test the above gauge/gravity duality. 
In field theory side, we calculate 
the vacuum expectation values (vevs) of the 
chiral primary operator (CPO) with conformal dimension 
$\Delta=1$ for all possible supersymmetric vacua of the mABJM theory
with any $k$. 
In gravity side, we implement 
the Kaluza-Klein (KK) reduction on $S^7$ 
and construct 4-dimensional quadratic action 
from 11-dimensional supergravity 
on the AdS$_4\times S^7$ background. 
Applying holographic renormalizaton method~\cite{Henningson:1998gx},
we obtain an exact holographic relation 
for the vevs of the CPO with $\Delta=1$, 
which is given by $\langle {\cal O}^{(\Delta=1)}\rangle= N^{\frac32} \, f_{(\Delta=1)}$. 
Here we consider  $k=1$ case and $f_{(\Delta)}$ is 
a function of the conformal dimension and also depends on 
some parameters of LLM solutions~\cite{Lin:2004nb}. 
This result is extended to $k > 1$  for specific types of LLM solutions.

\section{Discrete Higgs Vacua and Dual Geometries}
%{\it Discrete Higgs Vacua and Dual Geometries.}---
The SU(4) global symmetry of the ABJM theory 
is broken to ${\rm SU(2)}
\times {\rm SU(2)}\times {\rm U}(1)$ symmetry under the supersymmetry 
preserving mass deformation~\cite{Hosomichi:2008jb,Gomis:2008vc}. 
To manifest the broken symmetry, 
we split the scalar fields into $Y^A = (Z^a, W^{\dagger a})$, 
where $A = 1,2,3,4$ and $a = 1,2$. 
Then the Higgs vacua of  the mABJM theory are
represented as direct sums of irreducible $n\times (n+1)$ 
GRVV matrices~\cite{Gomis:2008vc}, ${\cal M}_a^{(n)}$ with 
the occupation number $N_n$, and their Hermitian conjugates
$(n+1)\times n$ matrices
$\bar {\cal M}_a^{(n)}$ with $N_n'$~\cite{Cheon:2011gv}. 
(See also \cite{Kim:2014yca} for the vacuum solutions.) 
Since $Z^a$ and $W^{\dagger a}$ are $N\times N$ matrices, 
we have two constraints, $\sum_{n=0}^\infty\left(n + \frac12 \right)
\left(N_n + N_n'\right) = N$ and $\sum_{n=0}^\infty N_n 
= \sum_{n=0}^\infty N_n'$. 
In addition, in order to have supersymmetric vacua 
the range of the occupation numbers should be  
$0\le N_n,\, N_n' \le k$~\cite{Kim:2010mr,Cheon:2011gv}. 
As a result, the supersymmetric vacua of the mABJM theory 
are completely classified in terms of the occupation numbers, $\{ N_n,\, N_n'\}$.  

The LLM geometry with $\mathbb{Z}_k$ orbifold is 
determined by two functions 
$Z$ and $V$, 
\begin{align}\label{zV}
Z(\tilde x,\tilde y) &= \sum_{i=1}^{2N_{\rm B}  + 1}  \frac{(-1)^{i+1} (\tilde x-\tilde x_i)}{
 2\sqrt{ (\tilde x-\tilde x_i)^2 + \tilde y^2 }},
\nn \\
V(\tilde x,\tilde y) &= \sum_{i=1}^{2N_{\rm B} +1} \frac{(-1)^{i+1}}{
2 \sqrt{ (\tilde x-\tilde x_i)^2 + \tilde y^2 }},
\end{align}
where $\tilde{x}$ and $\tilde{y}$ are 11-dimensional coordinates, 
$\tilde x_i$'s the positions of boundaries of black and white strips
in the droplet picture~\cite{Lin:2004nb}, and $N_{\rm B} $ is the number of finite black droplets. 
Due to the quantization condition of the 4-form flux,  
the difference between consecutive $\tilde x_i$'s 
is quantized as 
$\tilde x_{i+1} - \tilde x_i = 2\pi l_{\rm P}^3 \mu_0 {\mathbb Z}$ 
with the Planck length $l_{\rm P}$ 
and the mass parameter $\mu_0$. 
This implies that all possible LLM geometries 
are parametrized by the quantized $\tilde x_i$'s. 
For the asymptotic expansion of the LLM geometries, 
it is convenient to introduce new parameters~\cite{Kim:2016dzw},
\begin{align}\label{Cp}
C_p = \sum_{i=1}^{2 N_{\rm B}  +1}(-1)^{i+1} \left(\frac{\tilde x_i}{2 \pi l_{\rm P}^3\mu_0 
\sqrt{A}}\right)^p,
\end{align}
where $A = k N -\frac12 \sum_{n=0}^{\infty} \left[l_n (k-l_n) + l_n' (k-l_n')\right]$ is the area of the Young diagram picture and 
$\{ l_n, \,l_n\}$ are set of parameters classifying the LLM geometries in the droplet picture. See \cite{Cheon:2011gv} for the details.
There is one-to-one correspondence between $\{l_n, l_n'\}$ and the occupation numbers 
$\{N_n, N_n'\}$ in the vacua of the mABJM theory~\cite{Cheon:2011gv}. 

\section{Kaluza-Klein Holography}
In order to implement the KK holography 
method~\cite{Skenderis:2006uy,Skenderis:2006di, Skenderis:2007yb}, 
we consider asymptotic expansion 
of the LLM geometries 
with $\mathbb{Z}_k$ orbifold 
and regard the deviation from 
${\rm AdS}_4\times S^7/{\mathbb Z}_k$ geometry 
as solutions to perturbed equations of motion  
in 11-dimensional supergravity on such background.
According to the dictionary of the gauge/gravity 
duality~\cite{Gubser:1998bc,Witten:1998qj}, 
the deviations in asymptotic limit encode 
the information of vevs of CPOs~\cite{Klebanov:1999tb} 
in the mABJM theory.

Our purpose in this letter is to compare quantitatively 
the vevs of CPOs in the mABJM theory with the corresponding 
asymptotic coefficients of 
KK scalar fields, based on the KK holographic 
procedure~\cite{Skenderis:2006uy,Skenderis:2006di,Skenderis:2007yb}. 
Since the elements of the GRVV matrices are 
real numbers, one can compute the vevs of CPOs  in terms of numerical values and compare them with the corresponding coefficients of the KK scalars in gravity side.
The number of supersymmetric vacua is numerous for a given $N$ 
and thus large number of nontrivial tests can be carried out.

More precisely, 
the vev of CPO with conformal dimension $\Delta$ 
is proportional to the coefficient of $z^\Delta$-term 
in the asymptotic expansion of the dual scalar field 
in gravity side~\cite{Skenderis:2006uy}, 
where $z$ represents the coordinate in holographic direction.  
When we restrict our interest to 
the CPOs with low conformal dimensions, 
it is sufficient to consider 
the dual LLM geometry near the asymptotic limit.

In particular, the vevs of CPO with $\Delta=1$ 
are holographically determined by the solutions 
of the  linearized supergravity equations of motion 
on the AdS$_4\times S^7/\mathbb{Z}_k$ background. 
In this case, diagonalized gauge invariant fields 
in 11-dimensions 
can be identified with 4-dimensional gravity fields 
without nontrivial field redefinitions. 
However, for $\Delta\ge 2$, 
nonlinear terms in the equations of motion are  
not negligible and nontrivial field redefinitions 
in the construction of the 4-dimensional gravity theory 
are necessary~\cite{Skenderis:2006uy}. 
In this letter we  focus on the CPO 
with $\Delta = 1$ 
and leave our study on 
CPOs with $\Delta\ge2$ as future work.

\subsection{Field theory side}
The CPO with conformal dimension $\Delta$ 
in the ABJM theory is
\begin{align}\label{CPOabjm}
{\cal O}^{(\Delta)}=C_{A_1,\cdots,A_{\Delta}}^{B_1,\cdots,B_{\Delta}}{\rm Tr}\big(Y^{A_1}Y^\dagger_{B_1}\cdots Y^{A_{\Delta}}Y^\dagger_{B_{\Delta}}\big),
\end{align}
where the coefficients, $C_{A_1,\cdots,A_{\Delta}}^{B_1,\cdots,B_{\Delta}}$, 
are symmetric in upper as well as lower indices and 
traceless over one upper and one lower indices.
The CPO in \eqref{CPOabjm} is written by reflecting
the global SU(4) symmetry of the ABJM theory. 
On the other hand, in the mABJM theory the CPO should reflect the SU(2)$\times$SU(2)$\times$U(1) 
global symmetry, of which the explicit form will be given later.

For a given vacuum, the complex scalar fields near the vacuum are written as  
$Y^A(x) = Y_0^A + \hat Y^A$, 
where $Y_0^A$'s $(A=1,2,3,4)$ are the vacuum solutions 
represented by GRVV matrices,  
and $\hat Y^A$'s are field operators.  
Then the vev of a CPO with  dimension 
$\Delta$ for a  specific vacuum in the mABJM theory is given by 
\begin{align}\label{vevCPO}
\langle {\cal O}^{(\Delta)}(Y^A)\rangle_{m} &= {\cal O}^{(\Delta)}(Y_0^A)+ \langle\delta  {\cal O}^{(\Delta)}(\hat Y^A)\rangle_{0}  + \frac1N{\rm -corrections},
\end{align}
where $\langle  \cdots \rangle_m$ and  $\langle  \cdots \rangle_0$ denote the vevs in the mABJM theory and the ABJM theory, respectively. The $\frac1N$-corrections come from the contributions of multi-trace terms~\cite{Witten:2001ua,Berkooz:2002ug,Gubser:2002vv}.  Here we note that quantum corrections of scalar fields are absent due to the high supersymmetry of the mABJM theory.  The second term in the above equation is a one point function in a conformal field theory and is vanishing. Therefore, in the large $N$ limit we have
\begin{align}\label{vevCPO1}
\langle {\cal O}^{(\Delta)}(Y^A)\rangle_{m} &= {\cal O}^{(\Delta)}(Y_0^A).
\end{align}
The vacua parametrized by the occupation numbers 
$\{N_n,\, N_n'\}$ of the GRVV matrices  
are composed of $N\times N$ matrices having numerical matrix components. 
Therefore, the resulting vev $\langle {\cal O}^{(\Delta)}\rangle_{0} $ is a numerical value  
for a given $N$. 
We compare the specific value of vev with the corresponding asymptotic coefficient in gravity side.

\subsection{ Gravity side}
 We start with $k=1$ case and write the fluctuations of 11-dimensional supergravity fields 
on the AdS$_4\times S^7$ background as 
\begin{align}\label{flucgC}
g_{pq}
=g^{0}_{pq} + h_{pq},
\quad 
F_{pqrs}
=F^{0}_{pqrs} + f_{pqrs},
\end{align}
where $g^{0}_{pq}$ and $F^{0}_{pqrs}$ represent 
the background geometry. 
To construct the 4-dimensional gravity theory, 
we implement KK reduction on $S^7$. 
This reduction involves the expansion of the fluctuations in \eqref{flucgC} 
in terms of $S^7$ spherical harmonics. 
The expansion is generally expressed in terms of scalar, vector, and tensor spherical harmonics. 
However, the dual gravity fields of the CPO with 
$\Delta=1$ are built purely from 
the coefficients of the scalar spherical harmonics. 
Here the truncated expansion 
involving only the scalar spherical harmonics is given,
\begin{align}\label{hpqexp}
&h_{\mu\nu}(x,y)= h^{I_1}_{\mu\nu}(x)Y^{I_1}(y),
\quad h_{\mu a}(x,y)=s^{I_1}_\mu(x)\nabla_{a}Y^{I_1}(y),
\nn\\
&h_{(ab)}(x,y)=s^{I_1}(x)\nabla_{(a}\nabla_{b)}Y^{I_1}(y),
\nn\\
&h^a_{~a}(x,y) =\phi^{I_1}(x)Y^{I_1}(y), 
\nn \\
&f_{\mu\nu\rho\sigma}(x,y)=
4\nabla_{[\mu} s_{\nu\rho\sigma]}^{I_1}(x)Y^{I_1}(y),
\nn\\
&f_{\mu\nu\rho a}(x,y)=
- s^{I_1}_{\mu\nu\rho}(x)\nabla_{a}Y^{I_1}(y),
\end{align}
where $I_1$ is non-negative integer,  
$x$ denotes the AdS${}_4$ coordinates,  $y$  the $S^7$ coordinates,  and 
we divide the 11-dimensional indices $p,q,\cdots$ into 
the indices of AdS$_4$, $\mu, \nu,\cdots$, and those of $S^7$, $a,b,\cdots$. 
The notation $(ab)$ is for symmetrized traceless combination, 
while  the notation $[ab\cdots]$ is for anti-symmetrization among the indices, $a,b,\cdots$. 
The scalar spherical harmonic $Y^{I_1}$ is determined by the eigenvalue equation,
\begin{align}\label{sheqY}
\left[\nabla_a\nabla^a + \frac{I_1(I_1+6)}{L^2}\right] Y^{I_1} = 0,
\end{align} 
where $L = (32\pi^2 k  N)^{1/6} l_{\rm P}$ is the radius of $S^7$.
The expansion \eqref{hpqexp} follows the convention of  
\cite{Kim:1985ez,Skenderis:2006uy}. 
See also \cite{Biran:1983iy} for the linearized equations of motion 
on AdS$_4\times S^7$ background in de Donder gauge.

Plugging \eqref{hpqexp} into the linearized equations of motion on the  
${\rm AdS}_4\times S^7$ background and collecting relevant equations of motion, 
we obtain two diagonalized equations for KK scalar fields in 4-dimensions 
(see \cite{Jang:2016aug} for details), 
\begin{align}\label{diagpp}
&
\left[\nabla_\mu\nabla^\mu-\frac{(I_1+6)(I_1+12)}{L^2}\right]\Phi^{I_1}(x) =0,
\nn \\
&\left[\nabla_\mu\nabla^\mu - \frac{I_1(I_1-6)}{L^2} \right] \Psi^{I_1}(x) =0,
\end{align}
where 
\begin{align}\label{scme1}
&\Phi^{I_1} = \frac{(I_1+7)}{14(I_1+3)}\left[18(I_1-1)\hat\phi^{I_1}
+7\hat\psi^{I_1}\right],
\nn\\
&\Psi^{I_1} = \frac{(I_1-1)}{14(I_1+3)}\left[-18(I_1+7)\hat\phi^{I_1} 
+7\hat\psi^{I_1}\right]
\end{align}
with gauge invariant combinations, 
\begin{align}\nn
\hat\phi^{I_1} &= \phi^{I_1} +\frac{I_1(I_1+6)}{L^2} s^{I_1}, 
\nn \\
\hat\psi^{I_1} &= 18 g_0^{\mu\nu} h_{\mu\nu}^{I_1}  -L\epsilon^{\mu\nu\rho\sigma}
\nabla_{\mu} s_{\nu\rho\sigma}^{I_1}.\nn
\end{align}
In the subsequent discussion, we will expand the LLM solution as in \eqref{flucgC} and read 
the corresponding values of $\Phi^{I_1}$ and $\Psi^{I_1}$.

\section{Exact Holography}
%{\it Exact Holography for Finite $N$.}---
In mABJM theory, the CPO defined in \eqref{CPOabjm} is constrained 
by the ${\rm SU(2)}\times {\rm SU(2)}\times {\rm U}(1)$ global symmetry. 
In particular for the $\Delta = 1$ case,  we have~\cite{Jang:2016aug}
\begin{align}\label{CPO1}
{\cal O}^{(1)} &= \frac{1}{2\sqrt{2}}{\rm Tr}
\left(Z^a Z_a^\dagger - W^{\dagger a} W_a\right),
\end{align}
where the overall numerical factor is determined 
by the normalization condition,  $C_{A_1,\cdots,A_{\Delta}}^{(I)B_1,\cdots,B_{\Delta}}C_{B_1,\cdots,B_{\Delta}}^{(J)A_1,\cdots,A_{\Delta}}+ (\textrm{c.c.})= \delta^{IJ}$. 
We have verified that all CPOs with $\Delta=1$ except for ${\cal O}^{(1)}$ 
in \eqref{CPO1} have vanishing vevs for all supersymmetric vacua 
of the mABJM theory.

Plugging \eqref{CPO1} into \eqref{vevCPO1} and expressing the vacuum solutions 
in terms of the GRVV matrices, we obtain 
\begin{align}\label{vevCPO2}
\langle {\cal O}^{(1)}\rangle_m = \frac{k \mu }{4\sqrt{2}\, \pi } 
\sum_{n=0}^{2 N_{\rm B}  +1}n(n+1)(N_n - N_n').
\end{align}
Here  $\mu$ is the mass parameter in the mABJM theory 
and has the relation $\mu = 4\mu_0$ with the mass parameter $\mu_0$ in the LLM geometries. 

The LLM geometries near the asymptotic limit can be regarded as 
AdS$_4\times S^7/\mathbb{Z}_k$ plus small fluctuations. 
Though the gauge conditions of the LLM solutions in 11-dimensional supergravity 
are not clear, the 4-dimensional fields $\Phi^{I_1}$ and $\Psi^{I_1}$
 in \eqref{scme1} are gauge invariant and can be read from the asymptotic expansions. 
According to the holographic dictionary,  asymptotic coefficients of 
$\Phi^{I_1}$ and $\Psi^{I_1}$  encode 
the vevs of the corresponding CPOs in the mABJM theory.

Warp factors in the LLM geometries~\cite{Lin:2004nb} are completely fixed by 
 $Z$ and $V$ in  \eqref{zV}, which are  functions of $\tilde x$ and $\tilde y$.  
To implement the holographic renormalization procedure~\cite{Henningson:1998gx},  
we should rewrite the LLM solution in terms of the Fefferman-Graham (FG) 
coordinate system, 
\begin{align}\label{FGcoord}
ds^2_{{\rm FG}} &= g_1(z,\tau)\left(-dt^2 + dw_1^2+ dw_2^2\right) + \frac{L^2}{4z^2} dz^2
\nn \\
&~~+ g_2(z,\tau) d\tau^2 + g_3(z,\tau) ds_{S^3}^2 + g_4(z,\tau) ds_{\tilde S^3}^2, 
\end{align}
where $z$ is the holographic direction and 
$\tau $ is one of the $S^7$ coordinates in the asymptotic limit. 
For a general droplet parametrized by the $C_i$'s in \eqref{Cp}, 
the asymptotic expansion of these warp factors gives 
\begin{align}\label{warpfactor}
g_1 &= \frac{L^2}{4z^2}\left[1 -  \frac{2\tau \beta_3}{3\sqrt{2}}\mu_0 z 
+ {\cal O}(\mu_0^2)\right],
\nn \\
g_2 &= \frac{L^2}{4 (1-\tau^2)} + {\cal O}(\mu_0^2), 
\nn \\
g_3 &= \frac{L^2(1+\tau)}{2}\left[1+ \frac{(1+\tau)\beta_3}{3\sqrt{2}}\mu_0 z + {\cal O}(\mu_0^2)\right], 
\nn \\
g_4 &= \frac{L^2(1-\tau)}{2}\left[1- \frac{(1-\tau)\beta_3}{3\sqrt{2}}\mu_0 z + {\cal O}(\mu_0^2)\right]
\end{align}
with $\beta_3 = 2 C_1^3 - 3 C_1 C_2 + C_3$. 
From the asymptotic expansion of the warp factors and 
a similar expansion for the 4-form flux~\cite{Jang:2016aug},  
one can read fluctuations in \eqref{flucgC}, which will later be used in the construction of the modes $\Phi^{I_1}$ and $\Psi^{I_1}$. 

We need to express the LLM geometries in terms of the spherical harmonics 
on $S^7$. Since the geometries have SO(4)$\times$SO(4)  isometry, they can be 
appropriately expressed in terms of the spherical harmonics having the same isometry. 
The scalar spherical harmonics on $S^7$ are defined by the eigenvalue equation \eqref{sheqY}. 
In $\mu_0 z\to 0$ limit the warp factors in \eqref{FGcoord} depend only on 
the $\tau$ coordinate, and thus the appropriate spherical harmonics are the solutions 
of \eqref{sheqY} which also depends only on $\tau$ coordinate.   
One obtains two kinds of such solutions represented by the 
hypergeometric function $_2F_1(a,b,c;\tau^2)$,
which correspond to those with $ I_1= 4i$ and 
$I_1= 4i +2$, ($i=0,1,2,\cdots$).  
First few nonvanishing $Y^{I_1}$'s are given by 
\begin{align}\label{lowerY}
 Y^0= 1,\,\,\, Y^2 = \frac{\tau}{2\sqrt{2}}, 
 \,\,\, Y^4 = \frac{1- 5\tau^2}{8\sqrt{10}},
 \,\,\, \cdots \,\,\, , 
\end{align}
where we used the normalization 
$\frac{3}{\pi^4}\int Y^{I_1}Y^{J_1} = \frac{3 I_1! \delta^{I_1 J_1}}{ 2^{I_1-1} (I_1 + 3)!}$.

According to the dictionary of gauge/gravity duality, 
the mass of scalar mode is related to the conformal 
dimension $\Delta$ of the corresponding operator. 
In AdS$_4$/CFT$_3$ correspondence the relation is
\begin{align}\label{ScalarMass}
m^2 L_{{\rm AdS}_4}^2 = \frac{m^2 L^2}{4} = \Delta (\Delta -3). 
\end{align} 
From \eqref{diagpp} we see that the masses of the 4-dimensional scalar modes have the form 
$m^2 = n(n-6)/L^2$ with $n= I_1 + 12$ for $\Phi^{I_1}$ and 
$n= I_1$ for $\Psi^{I_1}$,  respectively. 
So the relation \eqref{ScalarMass} is rewritten as 
$n(n-6) = 4\Delta (\Delta -3)$.
From this relation the scalar mode $\Phi^{I_1}$ satisfying
the relation $(I_1+12) (I_1+6) = 4\Delta ( \Delta-3)$
can not be the dual scalar field of the CPO with $\Delta = 1$. 
On the other hand, we notice that 
the field $\Psi^{I_1}$ satisfies the relation 
$I_1(I_1-6) = 4\Delta (\Delta -3)$, which implies $\Delta = \frac{I_1}{2}$. 
We naturally expect that the dual scalar field 
for the CPO with $\Delta =1$ in \eqref{CPO1} 
is nothing but $\Psi^{I_1}$ in \eqref{scme1} with $I_1=2$.

By writing the asymptotic expansion of the LLM geometries  \eqref{warpfactor} 
in terms of the scalar spherical harmonics \eqref{lowerY},   
we obtain the asymptotic behavior of the 4-dimensional scalar modes, $\Phi^{I_1}$ 
and $\Psi^{I_1}$ with $I_1=2$, 
\begin{align}
\Phi^{I_1 = 2} = {\cal O}(\mu_0^3), 
\quad
\Psi^{I_1 = 2} = -24 \beta_3 \mu_0 z + {\cal O}(\mu_0^3). 
\end{align} 
According to the holographic renormalization 
procedure for the scalar action 
on the AdS$_4$ background, we have
\begin{align}\label{dual_rel}
\langle {\cal O}^{(1)}\rangle_m = \frac{N^2}{\sqrt{\lambda}}{\mathbb N}\, \psi^{(1)} 
= -  \frac{24 N^2}{\sqrt{\lambda}}\, {\mathbb N} \, \beta_3 \mu_0,
\end{align}
where ${\mathbb N}$ is a numerical number depending on the normalization of the scalar $\Psi^{I_1=2}$, $\psi^{(1)}$ is the coefficient 
of the radial coordinate $z$ in the expansion of the scalar mode, and $\lambda$ is the 't Hooft coupling constant defined as $\lambda = N/k$ in ABJM theory.  In the case $k=1$, the overall normalization in \eqref{dual_rel} is reduced to $N^{\frac32}$. 
The $N^2/\sqrt{\lambda}$-dependence in the right-hand side 
of \eqref{vevCPO3} is a peculiar behavior of the normalization 
factor in holographic dual relation for the M2-brane theory~\cite{Klebanov:1996un,Aharony:2008ug,Drukker:2010nc}. 
By identifying the occupation number of vacua in the mABJM theory 
with the discrete torsion in the LLM geometries~\cite{Cheon:2011gv} , i.e., 
\begin{align}\label{NNn}
\{N_n, \, N_n'\}\,\, \Longleftrightarrow \,\,\{l_n, \, l_n'\},
\end{align}
the normalization factor ${\mathbb N}$ is fixed.

Comparing the values in \eqref{vevCPO2} in $k=1$ field theory 
with the corresponding values of $\beta_3$ in gravity side,  
we obtain an exact holographic relation, 
\begin{align}\label{vevCPO3}
\langle {\cal O}^{(1)}\rangle_m =\frac{N^{\frac32}\mu_0}{3\sqrt{2}\,\pi}\, \beta_3
\end{align}
with the numerical normalization factor ${\mathbb N} = -\frac{\sqrt{2} }{144\pi}$.
In Young-diagram picture of the LLM geometries, $\beta_3$ has no dependence on $N$ and 
depends only on the shape of Young-diagrams and is independent of the size of the diagram. 
To prove the holographic relation \eqref{vevCPO3}, 
we used the relations, \eqref{vevCPO2} and  \eqref{NNn}, and the following identity, 
\begin{align}\nn
\sum_{n=0}^{2 N_{\rm B}  +1}n(n+1)(l_n - l_n') 
= \frac13\left(2\tilde C_1^3 - 3 \tilde C_1\tilde C_2 + \tilde C_3\right),
\end{align}
where $\tilde C_p \equiv  A^{\frac{p}{2}} C_p$, $A$ being the area 
of the Young-diagram  in \eqref{Cp}. 
Interestingly, the relation \eqref{vevCPO3} in the leading $N$ behavior is exactly satisfied for all $N$ ($\ge 2$). Technical details for the proof will be reported in a separate paper~\cite{Jang:2016aug}.

We also obtain the normalization factor ${\mathbb N}$ for 
$k\ne 1$ with $N_{\rm B}  = 1$. 
In the Young-diagram picture, this   
corresponds to the rectangular-shaped diagrams. 
For this case the exact dual relation is given by 
\begin{align}\label{vevCPOnek}
\langle {\cal O}^{(1)}\rangle_m =\frac{N \sqrt{k\tilde N}\mu_0}{3\sqrt{2}\,\pi}\, \beta_3
= \frac{N \sqrt{N\tilde N}\mu_0}{3\sqrt{2}\,\pi\,\sqrt{ \lambda}}\, \beta_3,
\end{align}
where $\tilde N = A/k$ and $\lambda= N/k$ is 't Hooft coupling constant in the ABJM theory.   In the large $N$ limit, $\tilde N$ approaches $N$ and the overall factor $N^2/\sqrt{\lambda}$ in \eqref{dual_rel} appears. 
For $k=1$, the holographic relation \eqref{vevCPOnek} reduces to 
the result in  \eqref{vevCPO3}.

\section{Conclusion}
%{\it Conclusion.}---
In this letter, we carried out the KK reduction and the holographic renormalization 
procedure for the mABJM theory and the LLM geometry in 11-dimensional supergravity. 
By calculating the vevs of CPO with $\Delta=1$ in field theory side and 
the corresponding asymptotic coefficients in gravity side, 
we found a supporting evidence for an exact gauge/gravity duality 
with  $k=1$ in the large $N$ limit. 
We could test the duality 
since discrete Higgs vacua exist in the mABJM theory and they correspond one-to-one with the LLM geometries. 
We also extended the exact holographic relation to the case of  any $k$ 
for LLM geometries represented by rectangular-shaped Young-diagrams. 

It seems that the Higgs vacua of the mABJM theory are parametrized by the vevs of CPOs 
and those are nonrenormalizable due to the high supersymmetry. 
This is similar to 
the case of the Coulomb branch in large $N$ limit 
in ${\cal N}=4$ super Yang-Mills theory~\cite{Skenderis:2006uy,Skenderis:2006di}. 
Though our quantitative results for the gauge/gravity correspondence 
involve infinite examples,
we need to accumulate more analytic evidences for CPOs with $\Delta$ ($\ge 2$) and 
$k$ ($\ge 1$)  to define supersymmetric vacua. 
One should also test the dictionary of the gauge/gravity duality for one point functions 
of vector and tensor fields. 
For instance, it is important to verify that 
one point functions of the energy-momentum tensor vanish for all 
possible supersymmetric vacua,  since the mABJM theory is a supersymmetric theory. 
We leave these issues for future study. 

One necessary condition of the supergravity approximation in AdS/CFT correspondence 
is the large $N$ limit. It was reported recently that the dual gravity limit of the ABJM 
theory is broken down at the sub-leading order of $N$ 
due to one-loop quantum correction~\cite{Liu:2016dau}. 
Therefore, though our results suggest that the LLM geometries 
can be well-defined backgrounds  in the point of the gauge/gravity duality, fluctuations on those backgrounds can go beyond the supergravity approximation. 
We need more investigation in this direction.

 Recently, it was reported that the mABJM theory on $S^3$ has no gravity dual for the 
 mass parameter larger than a critical value~\cite{Nosaka:2016vqf} 
 (see also \cite{Anderson:2014hxa,Anderson:2015ioa,Nosaka:2015bhf}). 
 Though the setup is different from ours, 
 which is the mABJM theory is on R$^{2,1}$, 
 it is also intriguing to investigate the large mass region for our case.
 It seems promising to pursue this issue since the LLM geometries have no singularity over the whole transverse region. 
 \\

\section*{Acknowledgements}
We would like to thank Changrim Ahn, Loriano Bonora,  Kyung Kiu Kim, and Chanyong Park for helpful discussions. This work was supported 
by the National Research Foundation of Korea(NRF)
grant with the grant number NRF-2014R1A1A2057066  and NRF-2016R1D1A1B03931090 (Y.K.) 
and NRF-2014R1A1A2059761 (O.K.).

%%%%%%%%%%%%%%%%%%%%%%%%%%%%%%%%%%%%%%%%%%%%%%

\end{document}